\documentclass[twocolumn,PRL,showpacs]{revtex4}
\usepackage{graphicx}

\begin{document}

\title{Eight types of physical "arrows" distinguished by Newtonian space-time symmetry.}

\author{J. Hlinka}
\affiliation{Institute of Physics, Academy of Sciences of the Czech Republic, Prague, Czech Republic}

\date{\today}

\begin{abstract}

The paper draws the attention to the spatiotemporal symmetry of
various vector-like physical quantities. The symmetry is specified
by their invariance under the action of symmetry operations of the Opechowski nonrelativistic space-time rotation group $O(3)\bigotimes \{1, 1'\}= O'(3)$, where $1'$ is time-reversal operation.
 It is argued that along with the canonical polar vector, there are another 7 symmetrically distinct classes of stationary physical
 quantities,
 which can be --  and often are  -- denoted as standard three-components vectors, even though they do not transform as a static polar vector under all operations of $O'(3)$.
The octet of symmetrically distinct "directional quantities"
can be exemplified by: two kinds of polar
 vectors (electric dipole moment {\bf P} and magnetic toroidal moment {\bf T}),
 two kinds of axial vectors (magnetization {\bf M} and electric toroidal moment {\bf G}),
  two kinds of chiral "bi-directors" {\bf C} and {\bf F} (associated with the so-called true and false chirality, resp.) and
  still another two achiral "bi-directors" {\bf N} and {\bf L}, transforming as the nematic liquid crystal order parameter  and as
  the antiferromagnetic order parameter of the hematite crystal $\alpha$-Fe$_2$O$_3$, respectively.
\end{abstract}



\maketitle 

Physical quantities defined by a magnitude
and an oriented axis in 3D space are often
represented by three-component Euclidean vectors.
 Frequently, polar and axial (or pseudo-) vectors are distinguished, depending on whether they change their
sense or not, respectively, upon the operation of spatial
 inversion (parity operation ${\bar{1}}$).\cite{knihaBirs64,Grim94,Asch74,Kops06}
 For classification of temporal processes or magnetic phenomena of vectorial nature, the action of the time-inversion operator (${ 1'}$)
  can be used.
 For example, magnetization $\bf M$ and magnetic field vector $\bf H$ are "time-odd axial"
 vectors
 (preserved by ${ \bar{1}}$ operation but changing their sign under the ${ 1'}$ operation),
 electric polarization $\bf P$ or electric field $\bf E$ are "time-even polar" vectors,
 while other quantities like velocity {\bf v} or toroidal moment
  $\bf T$ are "time-odd polar" vectors.\cite{knihaBirs64,Asch74,Sirotin75,Grim94,Kops06,Dubo90}
The two inversion operations generate an Abelian group of 4
elements $\{1,{\bar{1}}, 1', {\bar{1}'}\}$ and 4 one-dimensional irreducible
representations; the symmetry operations this group allows to classify these
vectors into 4 categories (see Table\,\ref{intro}).\cite{knihaBirs64,Grim94,Asch74,Kops06}

\begin{table}
\begin{tabular}{r|r|r|r|l c}
1 & ${ \bar{1}}$ & ${ 1'}$ & ${ \bar{1}'}$ & vectorial quantity & symbol
 \\
 \hline
1& 1&$ 1$& $ 1$ & electric toroidal moment & \bf{G}
\\
1&-1&$1$& -1& electric
dipole moment & \bf{P}\\
1&1&-1& -1  & magnetic dipole moment  &\bf{M} \\
1&-1&-1&  1 & (magnetic) toroidal moment & \bf{T} \\
\end{tabular}
\caption{Action of space (${\bar{1}}$) and time (${ 1'}$) inversion operations on selected
examples of vectorial quantities: 1 stands for the
invariance, -1 stands for the sign reversal.\cite{Asch74,Dubo87,Litv07}} \label{intro}
\end{table}

The aim of this paper is to emphasize that there are another four
types of quantities, which are also defined by a magnitude, an
axis and a geometrical sign, and which are also often associated
by 3D  vectors, but which possess a different
spatio-temporal symmetry than the examples given in
Table\,\ref{intro}). We are going to specify here all 8 types of
"directional quantities" (i) by describing their transformation
properties under the action of the elements
 of the Opechowski general space-time rotation group $O(3)\bigotimes \{1,
 1'\}=O(3){\bf .
 1'}=
 O'(3)$, (ii) by enumerating the  associated limiting
 groups defining their symmetry
 invariance, (iii) by providing several examples to each case.
 We shall also briefly discuss possibilities and
 difficulties with introduction of formal algebraic manipulations.
 Simultaneous considerations about all 8 different types of such directional quantities
 can be useful in various areas of physics.

\begin{figure}
\includegraphics[width=6.00cm]{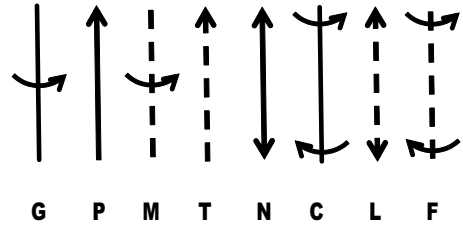}
 \caption{Pictographs of 8 kinds of
quantities defined by a sign, a magnitude and an axis. Letter
symbols allow to identify each pictograph with the symmetry
assignment given in Tables\,\ref{intro}-\ref{seznam}. Arrows in
pictographs drawn by dashed lines should be considered as
indicating a stationary current or motion (time inversion
operation does change their sense), while those in pictographs
drawn by full lines are time-irreversible (as e.g. the electric polarization). Pictographs were
inspired by pictures employed for similar purpose in
Refs.\,\onlinecite{Zheludevbook,Zhel86,Sirotin75}. }
\end{figure}

{\bf Basic symmetry argument.} These 8 symmetrically different species are resumed
pictographically in Fig.\,1. {\it Why do we have just 8 of such
quantities?} Let us consider any stationary physical quantity
${\bf X}$ (attached to a physical object),
 which simultaneously defines
 a two-valued, geometry-related sign, a nonnegative magnitude and a unique 1D
linear subspace of 3D Euclidean space (an axis of this quantity),
but nothing more. Since the quantity ${\bf X}$ defines a unique
axis in the space, the symmetry of ${\bf X}$ can be classified by
those $O(3). {\bf 1'}$ group operations that leave this axis
invariant. Such operations form an infinite subgroup of $O(3). {\bf 1'}$ that could be
denoted as the $\infty/m m . {\bf 1'}$ or $D_{\infty h}'$
 group.\cite{Dubo87,Schm08,Sirotin75}
  Moreover, it is natural to postulate that the magnitude of ${\bf X}$ ($|{\bf X}|\geq 0$)
    does not change under the operations of $O(3). {\bf 1'}$ group.
 This implies that transformation properties of ${\bf X}$ can be fully defined by specifying how its sign
  is
changed when elements of $\infty/m m . {\bf 1'}$ are applied to it.
  Since we restrict ourself only to the quantities for
which the sign of ${\bf X}$ can have only one of the two
possible values, the symmetry operation can either preserve the
sign or change it to the opposite one. In other words, the action of
the associated $\infty/m m . {\bf 1'}$ group operations consist in
multiplication of the geometrical sign of ${\bf X}$
either by 1 or by -1. In terms of theory of groups, this implies
that ${\bf X}$ transforms as one-dimensional (necessarily
irreducible) representation of the associated $\infty/m m . {\bf
1'}$ group. It is known that the $\infty/m m $ ($D_{\infty h}$) group has 4
distinct one-dimensional irreducible representations\cite{Hamermesh, Altmann}
so the $\infty/m m . {\bf 1'}$ ($D_{\infty h} \bigotimes {1,1'}$) one has
twice as much of them. Therefore, the physical quantities defined
by a sign, a magnitude and an axis can be classified in 8 symmetrically different
categories.

\begin{table}
\begin{tabular}{c|c|c|c|c|c|c|c|c|c}
irr. repr. & $E$ & $\bar{1}$ & $m_{\parallel}$ & $2_{\perp}$
 & $1'$ & $\bar{1}'$   & $m_{\parallel}'$ & $2_{\perp}'$ & symbol \\
 &  $\infty$ & $ \overline{\infty}$  &  &   & $\infty'$
  &  $\overline{\infty}' $  &  &   & \\
   &  $2_{\parallel}$ & $m_{\perp} $  &  &   & $ 2_{\parallel}'$
  &  $m_{\perp}' $  &  &   & \\
\hline
\bfseries $A_{1g}({\Sigma_{g}^{+}})  $ & 1 & 1 & 1 & 1 & 1 & 1 & 1 & 1 & {\bf N}\\
\bfseries $A_{2g}({\Sigma_{g}^{-}}) $ & 1 & 1 &-1 &-1 & 1 & 1 &-1
&-1 & {\bf G}\\
\bfseries $A_{1u}({\Sigma_{u}^{+}})  $ & 1 &-1 & 1 &-1 & 1 &-1 & 1 &-1 & {\bf P}\\
\bfseries $A_{2u}({\Sigma_{u}^{-}})  $ & 1 &-1 &-1 & 1 & 1 &-1 &-1 & 1 & {\bf C}\\
\bfseries $mA_{1g}({m\Sigma_{g}^{+}})$ & 1 & 1 & 1 & 1 &-1 &-1 &-1 &-1 & {\bf L}\\
\bfseries $mA_{2g}({m\Sigma_{g}^{-}})$ & 1 & 1 &-1 &-1 &-1 &-1 & 1 & 1 & {\bf M}\\
\bfseries $mA_{1u}({m\Sigma_{u}^{+}})$ & 1 &-1 & 1 &-1 &-1 & 1 &-1 & 1 & {\bf T}\\
\bfseries $mA_{2u}({m\Sigma_{u}^{-}})$ & 1 &-1 &-1 & 1 &-1 & 1 & 1 &-1 & {\bf F}\\
\end{tabular}
\caption{Characters of one-dimensional irreducible representations for selected elements of $\infty/m m . {\bf 1'} (
D'_{\infty h})$ group. The extra dash symbol identifies the operations
combined with time inversion. Irreducible representations are labeled similarly
as that of the $\infty/m m$ group\cite{Hamermesh,Altmann},
the "$m$" symbol indicates antisymmetry with respect  to the time inversion,
 similarly as it is adopted for grey\cite{Crac69}
  symmetry groups of magnetically ordered crystals\cite{Slaw12,Mato12,ISOTROPY}.
  }
  \label{irreps}
\end{table}

{\bf Classification by irreducible representations and basic examples.} 
The list of {\it all one-dimensional irreducible
representations} of the $\infty/m m . {\bf 1'}$ group is given in
Table\,\ref{irreps}. First column gives the irreducible
representations label following the convention used e.g. in
Refs.\,\onlinecite{Hamermesh} and \onlinecite{Altmann}, resp., the
last column contains a letter symbol used in Table\,\ref{seznam}.
and in Fig.\,1. Remaining columns in the table are associated with
the classes of symmetry elements of the $\infty/m m . {\bf 1'}$
  group. There are various
physical quantities having the listed transformation properties. For
example,  polarization ($\bf P$) and
magnetization ($\bf M$) transform as $A_{1u}({\Sigma_{u}^{+}}) $
and $mA_{2g}({m\Sigma_{g}^{-}})$ irreducible
 representations, resp. The symbol $\bf T$ invokes the often discussed  toroidisation  or toroidal
 moment,\cite{Eder07,Gorb94,Spal08,Dubo90,Kopa09}
 even though there are many other more frequently used quantities that also transform as the $mA_{1u}({m\Sigma_{u}^{+}})$
 representation, such as electric current, momentum or velocity of a particle, vector potential or the
 Poynting vector ${\bf S}={\bf E} \times {\bf H}$.
  It is clear from Table\,\ref{irreps} that this "magnetic" toroidal moment ${\bf T}$ has a different symmetry than
 the "electric" toroidal
 moment $\bf G$, the latter exploited  e.g. for characterization of  electric polarization vortex states of
 small ferroelectric particles\cite{Pros06,Pros09,Pros09b} or poloidal spin currents\cite{Gorb84}.
 Recently, spontaneous magnetic  toroidization has been found e.g. in
  Ba$_2$CoGe$_2$O$_7$ crystal\cite{Tole11},  the $\bf G$-type distortion
   has been identified e.g. in the "ferroaxial" structures of
   CaMn$_7$O$_{12}$ and
    RbFe(MoO$_4$)$_2$ crystals.\cite{John12,Most12,Hear12}

\begin{table}
\begin{tabular}{l|c|r|r|r| r}
& & ${ \bar{1}}$ & ${ 1'}$ & ~$m_{\parallel} $ & limiting group
\\
\hline
\bf{G}&time-even axial& 1&$ 1$& $ -1$ & $\infty/m . {\bf 1'}$  \\
\bf{P} & time-even polar&$-1$&$1$& 1  & $\infty m . {\bf 1'}$\\
\bf{M} &time-odd axial&1&$-1$& $ -1$  & $\infty/m m' $ \\
\bf{T}&time-odd polar&$-1$&$-1$& 1  & $\infty/m' m$  \\
\bf{N}&time-even neutral& 1&$ 1$& 1& $\infty/m m . {\bf 1'}$\\
\bf{C}&time-even chiral&$-1$&$1$& $ -1$ & $\infty2 . {\bf 1'}$\\
\bf{L}&time-odd neutral&1&$-1$& 1 & $\infty/mm  $\\
\bf{F}&time-odd chiral&$-1$&$-1$& $ -1$ & $\infty/m'm'  $ \\
\end{tabular}
\caption{List of 8 symmetrically distinct "arrow" quantities and
their transformation under three independent operations $\infty/m
m . {\bf 1'} ( D'_{\infty h} )$ group attached to the axis. ($m_{\parallel}$ stands for any mirror plane operation parallel to the axis.)}\label{seznam}
\end{table}

Let us note that $\bf G$ and $\bf M$ are symmetric with respect to
the perpendicular mirror plane operation $m_{\perp}$ and $\bf P$
and $\bf T$ are symmetric with respect to the parallel mirror
plane $m_{\parallel}$. Thus, none of these quantities is
chiral.\cite{knihaBarr04} In fact, only two irreducible
representations from the Table\,\ref{irreps}. fulfill the group
theoretical condition of a chiral object (absence of  improper
rotation symmetry, such as center of inversion or mirror
planes\cite{knihaBarr04}): $A_{2u}$ and $mA_{2u}$.
 They are naturally suitable for
representation of chiral directional quantities, as their
geometrical sign can reflect the sign of their enantiomorphism.
For example, a helix might be characterized by its axis, the
magnitude (given by the pitch of the helix) and a geometrical sign,
indicating whether the helix is right-handed or left-handed. Such
a chiral quantity {\bf C}  transforms as
$A_{2u}$ irreducible representation. As a beautiful example of $mA_{2u}$  quantity ($\bf
F$) can
be taken the  antiferromagnetic order parameter  of the  linear-magnetoelectric chromite crystal
Cr$_2$O$_5$.\cite{Kado04,Tole94}
 This latter kind of
chirality, reversible upon time reversal, is sometimes called
"false chirality".\cite{knihaBarr04,Barr00}

Finally, there are also two irreducible representations symmetric
with respect to both $m_{\parallel}$ and $m_{\perp}$ ($\bf L$ and
$\bf N$). The time-odd variant ($\bf L$) can be used to describe
another type of directional antiferromagnetic order parameter,
e.g. in the hematite crystal $\alpha$-Fe$_2$O$_3$.\cite{Tole94}
The fully symmetric ($A_{1g}$) representation is perhaps the most
singular one. It can be associated with the so-called director,
exploited in the theory of liquid crystals to characterize the
spontaneously parallel spatial orientation of rod-like molecules
in nematic phases.\cite{Lag07} In this particular case there is no
reason to define its geometrical sign. However, there are other
${\bf N}$-like quantities that do have a sign. For example, a
consistently defined Frank vector of a wedge
disclination\cite{Roma83,Kleinert,Fran58} should allow to
distinguish whether
  the disclination can be formed by removing or inserting material body adjacent to the plane of the cut.
At the same time, this disclination itself is invariant against
all  operations of the $\infty/m m . {\bf 1'} ( D'_{\infty h} )$ group.\cite{Roma83}


{\bf Classification in terms of symmetry invariance groups.} Table\,\ref{irreps} defines fully  transformation properties of various uniaxial
 quantities discussed above.
 For many purposes, it is enough to consider
only those symmetry operations, which leave the quantity
invariant.\cite{Dubo87,Schm08,Kris69}
Such operations form infinite subgroups of the $\infty/m m . {\bf 1'} $ group.
 They are listed for each irreducible representation in Table\,\ref{seznam}.
 The content of these invariance group can be easily figured out from the
 pictographic symbols in Fig.\,1.
In addition, each pictograph shows a segment indicating the
magnitude of the quantity and an arrow associated with its
geometric sign (see Figs.\,1 and 2). Arrows in pictographs drawn
by dashed lines should be considered as indicating a stationary
current or motion (time inversion operation does change their
sense). This is the case of {\it time-odd} quantities ({\bf L, M,
T, F}). In contrast, the arrows in pictographs drawn by full lines
should be considered as time-irreversible (time inversion
operation does not change them, as they have a
grey-group\cite{Crac69} symmetry). These pictographs stands for
the {\it time-even} quantities {\bf N, G, P, C}. Let us note that
the {\bf P, T, N, L} quantities, symmetric with respect to the
{\em parallel mirror} plane
 operation
 $m_{\parallel}$, have arrows only in the radial direction, while
 $m_{\parallel}$-antisymmetric quantities, {\bf G, M, C, F},   have all only tangential
 arrows (bend arrows should be understood as drawn on a visible curved surface of
a co-axial circular cylinder.) One can also easily distinguish the
single-arrow pictographs of $2_{\perp}$-antisymmetric quantities
{\bf G, P, M, T} ({\em proper vectors}) from all the double-arrow
graphical symbols standing for $2_{\perp}$-invariant quantities
 {\bf N, C, L, F}, which we  call here as {\em bi-directors}.

{\bf Meaning of the geometric parity signs, bi-directors.} 
The fact that the parity sign can be represented in this way
emphasizes its geometrical nature. Obviously, the {\it strict
meaning of the parity sign} relies on some convention, too. For
example, the vector of electric dipole moment is taken as pointing
towards the center of the positive charge (and not the opposite),
the arrow associated with the velocity of a particle is drawn
towards its future position (and not the opposite), the sense of
the electric current refers normally to the velocity of the
positive charges, and the arrow in the pictograph standing for
magnetic dipole moment is that of the equivalent positive
stationary electric current circulating around the indicated axis.

Another set of conventions is needed to facilitate the {\it
algebraic representation} of such quantities. Typically, a polar
vector is represented by three coordinates defined by its
scalar-product projections to an oriented set of three orthonormal
basis vectors. It is so practical that we tend to represent all
other quantities in a similar way.

 In case of "true" vector quantities (those of Table\,\ref{intro}),
such algebraic representation is  usually defined through the time
derivatives and vectorial products or equivalent rules. In fact, this representation
justify the common usage of the simple ${\bf P}$-arrow pictograph for
all other vector quantities of the Table\,\ref{intro}. For example, magnetic moment
${\bf m}$ of a current turn is defined as a vector perpendicular
to the turn and directed so that the current observed from the end
of vector ${\bf m}$ envelops the turn
counterclockwise.\cite{Zilb247} Therefore, the pictograph for ${\bf M}$ (as well as for ${\bf G}$ and ${\bf T}$) can be formally replaced by that of ${\bf P}$, even though these quantities actually do have a different symmetry (in fact, Fig.\,1 could conveniently serve as the replacement table).
Moreover, this algebraic representation allows to
calculate any scalar and vectorial products in the usual
 way. Interestingly, vectorial products of true vectors
are true vectors and scalar products of true vectors transforms as
one of the four possible scalar species\cite{Kops06} (time-even
scalar $\sigma$, time-even pseudoscalar $\epsilon$, time-odd
scalar $\tau$ and time-odd pseudoscalar $\mu$, see Table\,\ref{grupa11}).

\begin{table}
\begin{tabular}{l|r|r|l}
 & ${ \bar{1}}$ & {$ 1'$}  & examples\\
\hline
$\sigma$    & 1&$ 1$&   ${\bf G}.{\bf G}$, ${\bf T}.{\bf T}$,   ${\bf P}.{\bf P}$, ${\bf M}.{\bf M}$, $\nabla{\bf P}$, mass, charge         \\
$\epsilon$  &$-1$&$1$ &  ${\bf P}.{\bf G}$, ${\bf T}.{\bf M}$                        \\
$\tau$      & 1&$-1$&  ${\bf M}.{\bf G}$, ${\bf T}.{\bf P}$, time                                   \\
$\mu$       &$-1$&$-1$ & ${\bf T}.{\bf G}$, ${\bf M}.{\bf P}$, magnetic monopole     \\
\end{tabular}
\caption{Four scalar types\cite{Kops06} specified according to
their invariance under space-inversion and time-reversal
operations (time-even scalar $\sigma$, time-even pseudoscalar
$\epsilon$, time-odd scalar $\tau$ and time-odd pseudoscalar
$\mu$).} \label{grupa11}
\end{table}

 In case of bi-directors,  none of the SO(3) operations can reverse their geometrical sign.
 It indicates the fundamental difficulty with representation of
bi-directors by three-component algebraic vectors. In fact, each
of these bi-director quantities transforms as an "antitandem"
arrangement of two vectors - as a couple ("dipole") of two
opposite vectors ${\bf X}_1$ and ${\bf X}_2$ ($|{\bf X}_1|=|{\bf X}_2|$) arranged on a common
axis at some nonzero distance ${\bf r}_{21} = {\bf r}_{2} - {\bf
r}_{1}$. Obviously, ${\bf N}$ transforms as an antitandem of two
${\bf P}$ vectors, ${\bf C}$ as a antitandem of two ${\bf G}$
vectors, ${\bf L}$ as a ${\bf T}$ vector antitandem, ${\bf F}$ as
a ${\bf M}$ vector antitandem. Therefore, a bi-director can be
represented by a simple "two-body" term ${\bf a}_{12}={\bf X}_2
-{\bf X}_1$.
 Here it is assumed that the symmetry operations act both on the vectors and their position - operations that change ${\bf r}_{21}$ to the opposite are actually interchanging the sites 1 and 2.
The geometrical parity sign of such antitandem quantities could be
denoted as inward or outward, depending whether the vector ${\bf
X}_2$ is parallel or antiparallel to the vector ${\bf r}_{21}$,
and so its evaluation actually requires to know two quantities at
a time, ${\bf X}_2$ and ${\bf r}_{21}$. Having this in mind, a
range of algebraic operations can be
 nevertheless easily extended to all the above vectors and bi-directors.
 For the sake of convenience, types of the  quantities obtained as vectorial cross products or as multiplication by a scalar are given in
Table\,\ref{lookup}. Let us also note that from symmetry point of
view, time derivative acts here as multiplication by the time-odd
scalar $\tau$, so that e.g. the time derivative of the bi-director
${\bf L}$ transforms as the bi-director ${\bf N}$ and vice versa.

\begin{figure}
\includegraphics[width=6.00cm]{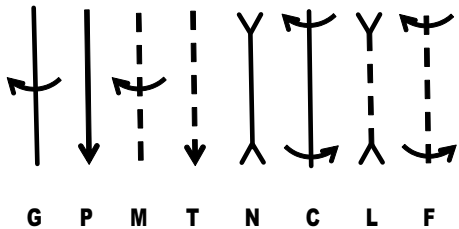}
 \caption{Pictographs of same 8 kinds of
quantities as in Fig.\,1, but with an opposite sense.}
\end{figure}

{\bf Classification of axes and concluding remarks.} 
In general, a physical object may have a physical property
transforming as one the 8 discussed cases only if its symmetry
invariance group is a subgroup of the corresponding limiting
group.
 For example,  macroscopic magnetization can exist only in crystals belonging to 31
 different Heesch-Shubnikov point groups,
  that are subgroups of $\infty/m m' $ group.\cite{Asch74,Opec86,Kris69}
If the axis of the limiting supergroup coincides with the symmetry
axis of the object, it is often named according to the associated
property (ferromagnetic axis, polar axis). Other axes could be
similarly labeled as toroidal, truly-chiral, falsely-chiral, ${\bf
G}$-axis, fully-symmetric and so on.

The term vector is sometimes employed to describe phenomena that
have a bi-director symmetry. For example, so-called Burgers vector
is widely used to characterize screw dislocations, which are
obviously non-polar, truly chiral ($C$-type) objects.
 Similarly, the antiferromagnetic vector\cite{Tole94, Kado04, Eder05}
 is often used to describe the  falsely-chiral ($F$-type)  antiferroelectric order.
  On the contrary, the so-called "chiral vector" or "vector chirality"\cite{Kawa88, Groh05, Mata11} is sometimes used to characterize cyclic spin arrangements on spin loops,
   for example in triangular antiferromagnetic lattices, even if the spin arrangement happens to have toroidal symmetry,
  which is "uni-directorial" but {\it achiral} (similarly as spin cycloids and N\'{e}el domain walls \cite{Tany11}).

\begin{table}
\vspace{1cm}
\begin{tabular}{lcl|c|c|c|c|c|c|c|c}
   &{\bf A} $\sim$&  &{\bf G} & {\bf P} & {\bf M} & {\bf T} & {\bf N} & {\bf C} & {\bf L} & {\bf F}\\
\hline
$ [{\bf G} \times {\bf A}]$ &or& $[\sigma {\bf A}] \sim$        &{\bf G} & {\bf P} & {\bf M} & {\bf T} & {\bf N} & {\bf C} & {\bf L} & {\bf F}\\
$ [{\bf P} \times {\bf A}]$ &or& $[\epsilon {\bf A}]\sim$       &{\bf P} & {\bf G} & {\bf T} & {\bf M} & {\bf C} & {\bf N} & {\bf F} & {\bf L}\\
$ [{\bf M} \times {\bf A}]$ &or& $[\tau   {\bf A}] \sim$        &{\bf M} & {\bf T} & {\bf G} & {\bf P} & {\bf L} & {\bf F} & {\bf N} & {\bf C}\\
$ [{\bf T} \times {\bf A}]$ &or& $[\mu   {\bf A}] \sim$         &{\bf T} & {\bf M} & {\bf P} & {\bf G} & {\bf F} & {\bf L} & {\bf C} & {\bf N}\\
\end{tabular}
\caption{Look-up table of transformation properties of vectorial
products and scalar multiplications. The symbol $\sim$ has a
meaning of "transforms as...", the operations involving
bi-director quantities ${\bf N}, {\bf C}, {\bf L}, {\bf F}$ are
defined in the text.}\label{lookup}
\end{table}

 Finally, it is well known that axial vector {\bf G} can be represented as a polar antisymmetric second-order tensor.
The bi-director quantities can be also classified within the
established tensorial
calculus.\cite{Litv94,knihaBirs64,Kop79,Kops06}
 They correspond to a special kind of second rank tensors,
 that has been once coined in  Russian literature as the "simplest tensor"
 (i.e. symmetrical second rank tensor having in its canonical form only a single nonzero element).\cite{Zheludevbook}
 In particular, {\bf N}-type bi-director could be considered as
 dual to the simplest time-even polar tensor,  {\bf C}-type bi-director  transforms as the simplest time-even axial tensor,
 {\bf L}-type bi-director  as the simplest time-odd polar tensor and {\bf F}-type bi-director
 as the simplest time-odd axial tensor.

 Nevertheless, we think that the unifying classification via irreducible
representations of $\infty/m . {\bf 1'}$ still provides a very
practical concept, applicable in various areas of physics.
 In solid state physics at least, the simple perspective where vectors and bi-directors have equal legitimacy
 might be useful when dealing with problems where several such quantities  are interacting,
  for classification of long-wavelength
 excitations or
 structural components of magnetoelectric multiferroic  crystals\cite{Schm08,Fieb05,Saxe11,Harr07}, for description of macroscopic properties of chiral objects,\cite{Jung89,Gopa11}             or for classification of topological defects like domain walls, magnetic vortices or skyrmions.\cite{Tany12b}
 In fact, we would like to offer a more complete discussion of possible applications of this concept in future
  and so we would be grateful to learn about other cases where this perspective  could bring some useful insight.

\section*{Acknowledgments}
It is a real pleasure to acknowledge V\'{a}clav Janovec for
his valuable suggestions related to this work, financially supported by the Czech Science Foundation (Project GACR 13-15110S).


\begin{thebibliography}{99}


 \bibitem{knihaBirs64} R. R. Birss, Symmetry and Magnetism (North-Holland, Amsterdam, 1964).

 \bibitem{Grim94} H. Grimmer, Ferroelectrics {\bf 161}, 181 (1994).


   \bibitem{Asch74} E. Ascher, Int. J. Magnetism {\bf 5}, 287 (1974).

  \bibitem{Kops06} V. Kopsk\'{y}, Z. Krystallogr. {\bf 221}, 51 (2006).

  \bibitem{Sirotin75} Yu. Sirotin and M. P. Shaskolskaya,
 Osnovy Kristallophiziki (In Russian, Moscow: Nauka, 1975);
[Engl. transl: Fundamentals of Crystal Physics (Moscow:
Mir. 1982)].

\bibitem{Hamermesh} M. Hamermesh, {\it Group Theory
 And its Applications to Physical Problems}
(Addison-Wesley, Reading, MA, 1964).

 \bibitem{Litv07} D. B. Litvin, Acta Cryst. {\bf A64}, 316 (2007).

\bibitem{Zheludevbook} I. S. Zheludev,  {\it Fizika kristallov i simmetria} (Nauka, Moscow, 1987).

\bibitem{Zhel86} I. S. Zheludev, Acta Cryst. {\bf A42}, 122 (1986).


\bibitem{Dubo87} V. M. Dubovik, S. S. Krotov, and V. V. Tugushev, Kristallografiya {\bf 32}, 540 (1987).

 \bibitem{Schm08} H. Schmid, J. Phys.: Condens. Matter {\bf 20}, 434201 (2008).



\bibitem{Altmann} S. L. Altmann and P. Herzig, {\it Point-Group Theory Tables}
(Oxford University Press, New York 1994).

 \bibitem{Crac69} A. P. Cracknell, Rep. Prog. Phys. {\bf 32}, 633 (1969).

\bibitem{Slaw12} W. Slawinski, R. Przenioslo, I. Sosnowska and V.
Petricek, Acta Cryst. B {\bf 68}, 240 (2012).

\bibitem{Mato12} J. M. Perez-Mato, J. L. Ribeiro, V. Petricek and M. I. Aroyo, J. Phys.:
Condens. Matter {\bf 24} 163201 (2012).

\bibitem{ISOTROPY} H. T. Stokes, D. M. Hatch, and B. J. Campbell, ISOTROPY program package Version 9.0, Department of
Physics and Astronomy, Bringham Young University, Provo, USA
(2007).


 \bibitem{Eder07} C. Ederer and N. A. Spaldin, Phys. Rev. B {\bf 76}, 214404 (2007).


 \bibitem{Gorb94} A. A. Gorbatsevich and Yu. V. Kopaev, Ferroelectrics {\bf 161}, 321 (1994).


 \bibitem{Spal08} N. A. Spaldin, M. Fiebig, and M. Mostovoy, J. Phys.: Condens. Matter {\bf 20}, 434203 (2008).

\bibitem{Dubo90} V. M. Dubovik and V. V. Tugushev, Physics Reports {\bf 187}, 145 (1990).


 \bibitem{Kopa09} Yu. V. Kopaev, Physics-Uspekhi {\bf 52}, 1111 (2009).


  \bibitem{Pros06} S. Prosandeev, I. Ponomareva, I. Kornev, I. Naumov, and L. Bellaiche, Phys. Rev. Lett. {\bf 96}, 237601 (2006).

  \bibitem{Pros09} S. Prosandeev, A. R. Akbarzadeh and L. Bellaiche, Phys. Rev. Lett. {\bf 102}, 257601 (2009).

  \bibitem{Pros09b} S. Prosandeev and L. Bellaiche, J. Mater. Sci. {\bf 44}, 5235 (2009).

\bibitem{Gorb84} A. A. Gorbatsevich and Yu. V. Kopaev, JETP {\bf 39}, 684
(1984).

\bibitem{Tole11} P. Toledano, D. D. Khalyavin and L. C. Chapon, Phys. Rev. B {\bf 84}, 094421 (2011).
\bibitem{John12} R. D. Johnson, L. C. Chapon, D. D. Khalyavin, P. Manuel,
 P. G. Radaelli, and C. Martin, Phys. Rev. Lett. {\bf 108}, 067201 (2012).
\bibitem{Most12} M. Mostovoy, Physics {\bf 5}, 16 (2012).
\bibitem{Hear12} A. J. Hearmon, F. Fabrizi, L. C. Chapon, R. D. Johnson, D. Prabhakaran, S. V. Streltsov, P. J. Brown, and P. G. Radaelli,
 Phys. Rev. Lett. {\bf 108}, 237201 (2012).


 \bibitem{knihaBarr04} L. D. Barron, {\it Molecular Light Scattering and Optical Activity} (Cambridge University Press, New York, 2004).


 \bibitem{Kado04} A. M. Kadomtseva, A. K. Zvezdin, Yu. F. Popov, A. P. Pyatakov, and G. P. Vorob'ev, JETP Letters {\bf 79}, 571 (2004).

\bibitem{Tole94} P. Tol\'{e}dano, Ferroelectrics {\bf 161}, 257 (1994).


 \bibitem{Eder05} C. Ederer and N. A. Spaldin, Phys. Rev. B {\bf 71}, 060401 (2005).

 \bibitem{Barr00} L. D. Barron, Nature {\bf 405}, 895 (2000).


 \bibitem{Lag07}  S. T. Lagerwall, {\it Ferroelectric and Antiferroelectric Liquid
 Crystals} (Wiley-VCH, 1999).


\bibitem{Roma83} A. E. Romanov and V. I. Vladimirov, Phys. Stat. Sol. (a) {\bf 78}, 11 (1983).

\bibitem{Kleinert} H. Kleinert,  {\it Multivalued Fields in Condensed Matter,
 Electromagnetism, and Gravitation} (World Scientific, Singapore, 2008), Chap.\,9.

 \bibitem{Fran58} F. C. Frank, Discuss. Faraday Soc. {\bf 5}, 19 (1958).

 \bibitem{Kris69} T. S. G. Krishnamurty and P. Gopalakrishnamurty, Acta Cryst. {\bf A25}, 333 (1969).

\bibitem{Zilb247} G. E. Zilberman, {\it Electricity and Magnetism},
(Moscow: Mir Publ., 1973).

\bibitem{Opec86} W. Opechowski, {\it Crystallographic and Metacrystallographic Groups} (North-Holland, Amsterdam, 1986).


\bibitem{Kawa88} H. Kawamura, J. Appl. Phys. {\bf 63}, 3086 (1988).


\bibitem{Groh05} D. Grohol, K. Matan, J.-H. Cho, S.-H. Lee, J. W. Lynn, D. G. Nocera, and Y. S. Lee, Nature Materials {\bf 4}, 323 (2005).


\bibitem{Mata11} K. Matan, B. M. Bartlett, J. S. Helton, V. Sikolenko, S. Matas, K. Prokes, Y. Chen, J. W. Lynn, D. Grohol, T. J. Sato, M. Tokunaga, D. G. Nocera, and Y. S. Lee, Phys. Rev. B {\bf 83}, 214406 (2011).


\bibitem{Tany11} B. M. Tanygin, Journal of Magnetism and Magnetic Materials {\bf 323}, 616 (2011).


 \bibitem{Litv94} D. B. Litvin, Acta Cryst. {\bf A50}, 406 (1994).



 \bibitem{Kop79} V. Kopsk\'{y}, Acta Cryst. {\bf A35}, 95 (1979).



 \bibitem{Fieb05} M. Fiebig, J. Phys. D: Appl. Phys. {\bf 38}, R123 (2005).

 \bibitem{Saxe11} A. Saxena and T. Lookman, Phase Transitions {\bf 84}, 421 (2011).

  \bibitem{Harr07}  A. B. Harris, Phys. Rev. B {\bf 76}, 054447 (2007).


 \bibitem{Jung89} P. Jungwirth, L. Sk\'{a}la, and R. Zahradn\'{\i}k, Chemical Physics Letters {\bf 161}, 502 (1989).

 \bibitem{Gopa11} V. Gopalan and D. B. Litvin, Nature Materials {\bf 10}, 376 (2011).

\bibitem{Tany12b} B. M. Tanygin, Physica B {\bf 407}, 866 (2012).


 \end{thebibliography}
 \end{document}